\documentclass[11pt]{article}
\usepackage{amssymb}
\usepackage{graphics}
\usepackage{epsfig}
\usepackage{a4wide}

\begin{document}

\title{ Revisiting the QCD Corrections to the R-Parity Violating Processes
$p\bar{p}/pp \to e\mu+X$ } \vspace{3mm}

\author{{ Wang Shao-Ming, Han Liang, Ma Wen-Gan, Zhang Ren-You, and Jiang Yi }\\
{\small Department of Modern Physics, University of Science and Technology of China (USTC),} \\
{\small Hefei, Anhui 230026, People's Republic of China} }

\date{}
\maketitle \vskip 12mm

\begin{abstract}
We present the theoretical predictions up to QCD next-to-leading
order for the cross section of high-mass electron-muon pair
production at the Tevatron and at the Large Hadron Collider(LHC),
considering only the dominant contributions from the
third-generation sneutrino. The dependence of the renormalization
and factorization scales on the total cross section, and the
effects on the $K$-factor due to the uncertainty of parton
distribution function(PDF) are carefully investigated. By
considering soft-gluon resummation effects to all orders in
$\alpha_s$ of leading logarithm, we present the transverse
momentum distributions of the final $e\mu$ pair.
\par
\end{abstract}

\vskip 5cm


\vfill \eject

\baselineskip=0.32in

\renewcommand{\theequation}{\arabic{section}.\arabic{equation}}
\renewcommand{\thesection}{\Roman{section}.}
\newcommand{\nb}{\nonumber}

\makeatletter      
\@addtoreset{equation}{section}
\makeatother       

\par
\begin{figure}[htp]
\scalebox{0.8}[0.8]{\includegraphics*[146pt,580pt][445pt,680pt]{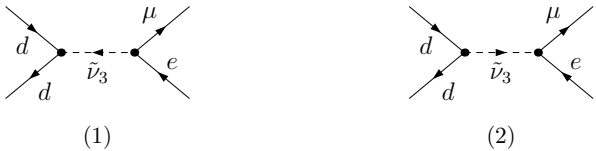}}
\caption{\label{fig1} Tree-level Feynman diagrams for subprocess
$d\bar d \to e\mu$. }
\end{figure}
\begin{figure}[htp]
\scalebox{0.8}[0.8]{\includegraphics*[125pt,406pt][560pt,655pt]{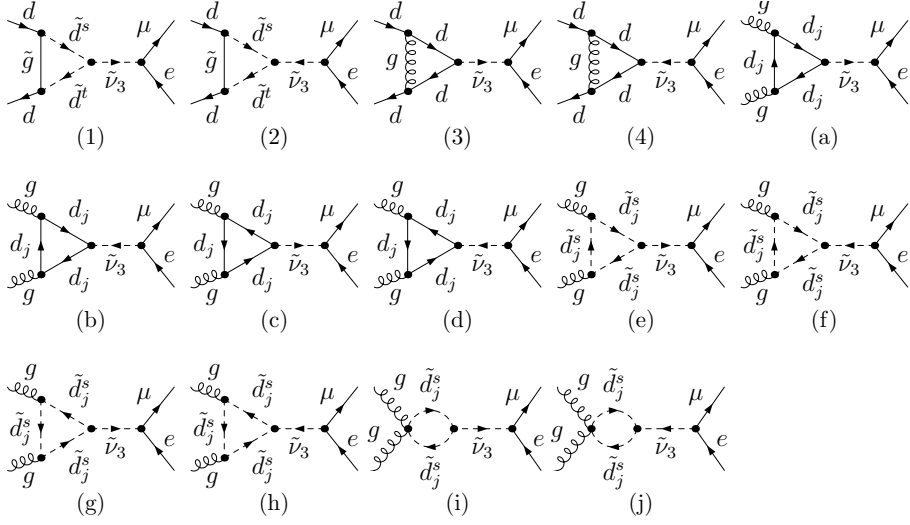}}
\caption{\label{fig2} Fig.2. QCD one-loop diagrams (1)-(4) for
subprocess $d\bar d \to e\mu$. QCD one-loop diagrams (a)-(j) for
$gg \to e\mu$, where the superscripts $s,t(=1,2)$ represent two
physical scalar quarks and the subscript $j(=1,2,3)$ is for three
generations. }
\end{figure}

Observation of electron-muon resonance high invariant mass ($Q$)
at hadron colliders could provide evidence of R-parity
violating(RPV) interactions. The $e\mu$ pair productions at hadron
colliders induced by RPV interactions at the leading order were
investigated in Ref.\cite{ppem}. In Ref.\cite{ind,deu,chen} the
QCD next-to-leading order(NLO) corrections to resonant sneutrino
production at hadron colliders were studied, while the QCD
corrections to $R$-violating process $p\bar p/pp \to q\bar q, gg
\to e\mu+X$ involving three generations of sneutrinos and squarks
was discussed in Ref.\cite{Wang}. Since there are increasing
interests in searching for high-mass $e\mu$ resonance at hadron
colliders \cite{CDF}, we revisit this topic to provide thorough
theoretical prediction as a reference for experimental analysis.
In this Letter only resonance contributions from the
third-generation sneutrino are involved in high-mass $e\mu$ search
under the single dominance assumption, \cite{Rept} and the
contributions from squark-exchanging diagrams are neglected by
applying a high threshold cut $(Q_0)$ on $e\mu$ invariant
mass.\cite{ppem,Wang} The tree-level and the one-loop QCD NLO
diagrams for sneutrino $e\mu$ resonance subprocess considering in
our calculation are depicted in Figs.\ref{fig1} and \ref{fig2},
respectively.

\par
We adopt the dimensional regularization(DR) method and the
modified minimal subtraction ($\overline{\rm MS}$) scheme. After
renormalization procedure, the virtual correction part of the
cross section is UV-finite. The IR divergences from the one-loop
diagrams will be cancelled by adding the soft real
gluon/light-quark emission corrections by using the two cutoff
phase space slicing method (TCPSS).\cite{twocut} The remaining
collinear divergences can be absorbed into the parton distribution
functions(PDF).

\par
We use the CTEQ6L parton distribution functions for the tree-level
cross sections and CTEQ6.1M for the QCD NLO corrected
ones.\cite{pdfs,pdf40} During the numerical calculation, we also
investigate the uncertainty induced by the factorization scale
$\mu_f$ and the CTEQ6 PDF. We take $40$ sets of CTEQ61.xx
PDF's\cite{pdf40}(set number goes form 201 to 240) to estimate the
uncertainty induced by the PDF. Actually, in the precise
calculation of the distributions of the transverse momentum($q_T$)
for the $e\mu$ pair, the quantitative comparison of $q_T$ and $Q$
is very crucial. When the $q_T$ value is comparable with $Q$ or
larger, fixed order perturbation theory gives sufficiently
accurate results. However, when $q_T \ll Q$, large logarithmic
terms, such as $\left[\alpha_s \ln \left(q_T / Q\right)\right]^n$,
arise at fixed order perturbation calculations and need to be
resummated. Therefore, we adopt the standard procedure\cite{resum}
to resummate the multiple soft gluon effects on $q_T$
distribution.

\par
The R-violating lagrangian relevant to present discussion is
expressed as\cite{potential1}
\begin{eqnarray}
\label{lag} {\cal L}_{\rlap/R}&=& \frac{1}{2} \lambda_{ijk}\cdot
(\bar \nu_{Li}^c e_{Lj} \tilde e_{jL}^* + e_{Li} \bar \nu_{Lj}^c
\tilde e_{Rk}^* + \nu_{Li} e_{Lj} \bar e_{Rk} - e_{Li} \tilde
\nu_{Lj} \bar e_{Rk}) + \nb\\
&&~~\lambda_{ijk}^{'}\cdot(\bar \nu_{Li}^c d_{Lj} \tilde d_{Rk}^*
- e_{Ri}^c u_{Lj} \tilde d_{Rk}^* + \nu_{Li} \tilde d_{Lj} \bar
d_{Rk} - e_{Li} \tilde u_{Lj} \bar
d_{Rk} +\nb\\
&&~~~~~~~~~~\tilde \nu_{Li} d_{Lj} \bar d_{Rk} - \tilde e_{Li}
u_{Lj} \bar d_{Rk})~+~h.c.
\end{eqnarray}
where $i,j,k$ = 1,2,3 are generation indices, the superscript $c$
refers to charge conjugation, $\lambda$ and $\lambda^{\prime}$ are
dimensionless R-violating Yukawa couplings, and $\lambda$ behaves
as $\lambda_{ijk}=-\lambda_{jik}$.

\par
In the numerical calculations, we take the RPV parameters
$\lambda$ and $\lambda^{'}$ to be real for simplicity with the
values as: $\lambda_{312}=0.062$, $\lambda_{321}=0.070$,
$\lambda'_{311}=0.11$, which are under the experimental
constraints presented in Ref.\cite{Rept}. We set the factorization
and the renormalization scales being equal and $\mu_f = \mu_r =
m_{\tilde\nu}$. The invariant mass cut of the $e\mu$ pair is set
to be $Q_0 = 50$ GeV. We apply the naive fixed-width scheme in the
sneutrino propagator to avoid the possible resonant
singularities(here we fix $\Gamma_{\tilde{\nu}}=10$ GeV as
demonstration). In principle, the value choice of the width of
sneutrino has an influence on the cross section, but does not
affect the $K$-factor. Since the sneutrino is non-colored
supersymmetric particle, there is no problem with double counting
in the QCD NLO calculation of the $d \bar d \to e\mu$ cross
section. The gluino and squark masses are taken as $m_{\tilde
g}=m_{\tilde q}=1~TeV$, in order to decouple the interactions
involving gluino and squarks and neglect the contributions of
squark-exchange diagrams. We have verified that the total cross
section involving the QCD NLO corrections is independent of the
cutoffs $\delta_{s}$ and $\delta_{c}$ in adopting the TCPSS
method. In the following calculation, we fix the soft cutoff as
$\delta_s=10^{-3}$ and collinear cutoff as $\delta_c=\delta_s/50$.
The calculations are carried out at the Tevatron and the CERN LHC
with $p\bar{p}$ colliding energy $\sqrt{s} = 1.96~TeV$ and $pp$
colliding energy $\sqrt{s}=14~TeV$, respectively. Since the
$\overline{MS}$ scheme violates supersymmetry, the
$q\tilde{q}\tilde{g}$ Yukawa coupling constant $\hat{g}_s$ takes a
finite shift at one-loop order as\cite{shiftgs}: $\hat{g}_s = g_s
\left [1 +\frac{\alpha_s}{8\pi}\left (\frac{4}{3}N_c - C_F\right
)\right ]$, with $N_c=3$ and $C_F=4/3$. We shall take this
coupling strength shift between $\hat{g}_s$ and $g_s$ into account
in our calculation.

\par
In Fig.3(a) we depict the curves of the tree-level and QCD NLO
corrected cross sections($\sigma^{0}$ and $\sigma^{QCD}$) of the
processes $p\bar{p}/pp \to e^+\mu^-+X$ versus the sneutrino mass
$m_{\tilde{\nu}}$ at the Tevatron and the LHC. Their corresponding
$K$-factors($K \equiv \frac{\sigma^{QCD}}{\sigma^{0}}$) as a
function of $m_{\tilde{\nu}}$ are depicted in Fig.3(b). We can see
that both the $K$-factor curves for the Tevatron and the LHC
colliders in Fig.3(b) show the difference between the curve
tendencies of $K$-factors for processes $p\bar p(pp) \to e \mu+X$
and $p\bar p(pp) \to \tilde{\nu}+X$. For the later process, both
the calculations in Ref.\cite{ind} and our cross-check for
confidence show that the $K$-factor curve for the Tevatron always
goes down when $m_{\tilde{\nu}}$ varies from $200$ GeV to $1~TeV$,
while the $K$-factor curve for the LHC goes up with the increment
of $m_{\tilde{\nu}}$ from $100$ GeV to $600$ GeV. It manifests
that the QCD NLO corrections to high-mass $e\mu$ resonance
production at both the Tevatron and the LHC cannot be adopted
directly from those for the single $\tilde{\nu}$ production
process as presented in Refs.\cite{ind,chen}. We can read out from
Fig.3(b) that the $K$-factors vary in the ranges of
$[1.182,~1.643]$ at the Tevatron and $[1.335,1.614]$ at the LHC.
\begin{figure}[htp]
\includegraphics[scale=0.7, bb=0 0 280 240]{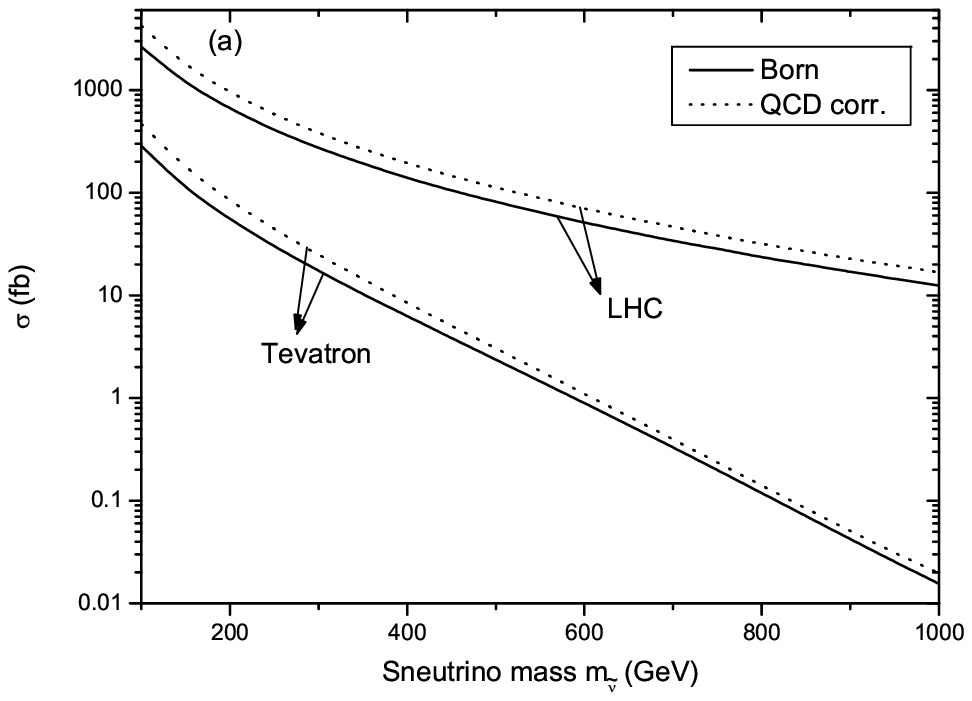}
\includegraphics[scale=0.7, bb=0 0 280 240]{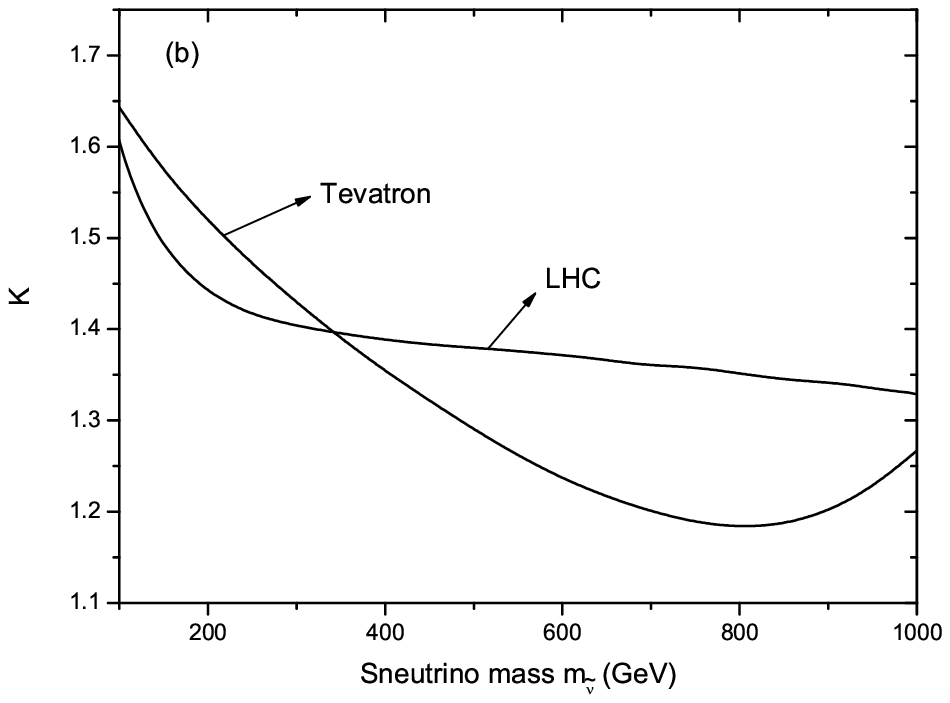}
\caption{The tree-level and total QCD NLO corrected cross sections
of the processes $p\bar{p}/pp \to e\mu+X$ at the Tevatron and the
LHC as a function of the sneutrino mass $m_{\tilde{\nu}}$ are
shown in Figure 3(a). Figure 3(b) shows the corresponding
relations between the $K$-factors and the sneutrino mass
$m_{\tilde{\nu}}$. }
\end{figure}

\par
Figures 4(a) and 4(b) demonstrate the dependence of $K$-factor on
the factorization scale $\mu_{f}/m_{\tilde{\nu}}$, when the
sneutrino mass is set to be $m_{\tilde{\nu}} = 100$, $250$, $500$
GeV. From the two figures we can estimate the uncertainty of the
QCD NLO correction induced by scale parameter $\mu_f$. In
Fig.4(a), we can read out that in the scale $\mu/m_{\tilde{\nu}}$
region of $[0.5,2]$ the $K$-factors at the Tevatron vary in the
ranges of $[1.639,~1.645]$, $[1.446,~1.498]$ and $[1.251,~1.328]$
corresponding to $m_{\tilde{\nu}} = 100$, $250$ and $500$ GeV
respectively. Figure 4(b) shows that the $K$-factors at the LHC
are in the ranges of $[1.567,~1.668]$, $[1.396,~1.434]$ and
$[1.362,~1.375]$ in the scale region of $\mu/m_{\tilde{\nu}}\in
[0.5,2]$ for $m_{\tilde{\nu}} = 100$, $250$ and $500$ GeV
separately. From Figs.4(a) and 4(b) we can see the relative errors
of $K$-factor induced by the factorization scale $\mu_f$ for
$m_{\tilde{\nu}}=100$ GeV, $250$ GeV, $500$ GeV in the scale
region $\mu_f/m_{\tilde{\nu}}\in [0.5,~2]$ are $0.17\%(3.1\%)$,
$1.8\%(1.3\%)$ and $3.0\%(0.46\%)$ at the Tevatron(LHC),
respectively.
\begin{figure}[htp]
\includegraphics[scale=0.7, bb=0 0 280 240]{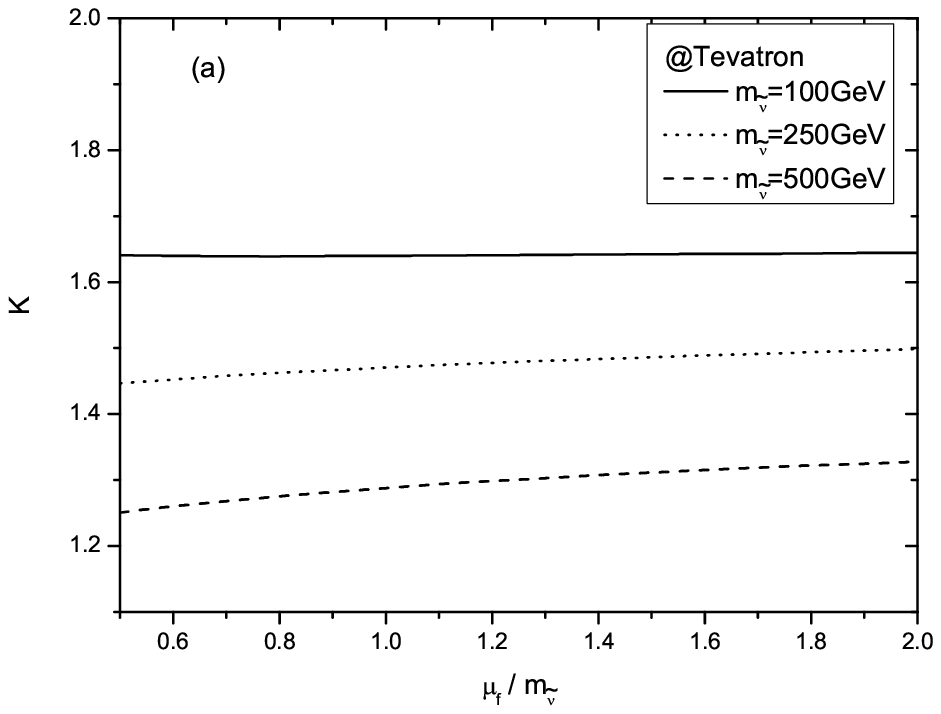}
\includegraphics[scale=0.7, bb=0 0 280 240]{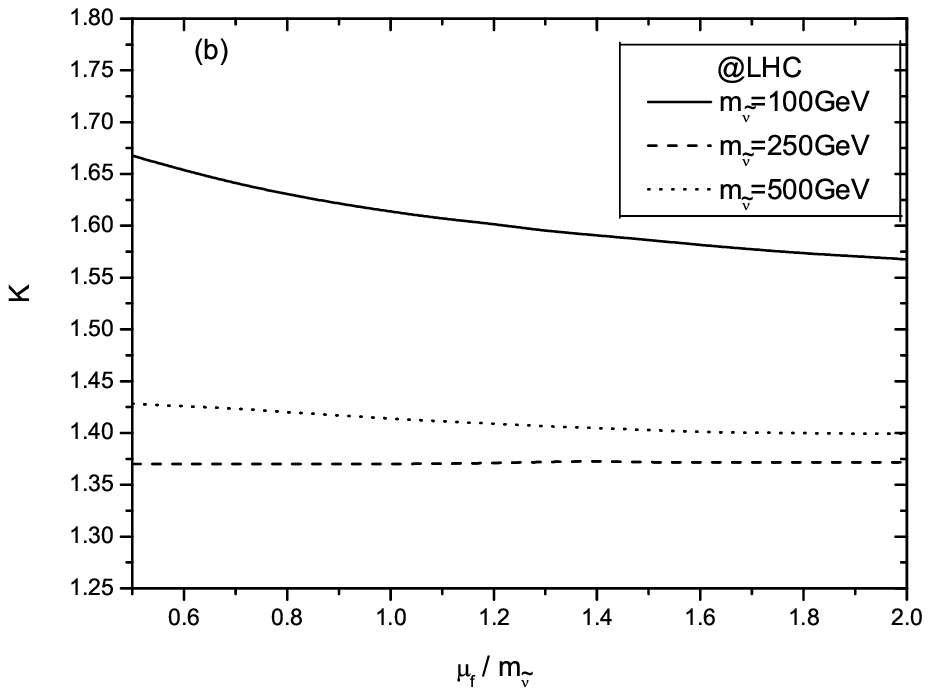}
\caption{Dependence of $K$-factor on the factorization scale
$\mu_{f}/m_{\tilde{\nu}}$. (a) at the Tevatron, (b) at the LHC. }
\end{figure}

\par
We investigate the uncertainty range due to the different CTEQ
sets. In Table 1 we list the $K$-factor values obtained by using
different CTEQ61.xx PDF sets, where the $K$-factor obtained from
the best fit CTEQ6.1M PDF is taken as the central value at
sneutrino mass. From the data in Table 1 we find that the
deviations of $K$-factor from the central value at the Tevatron
are in the ranges of $[-0.053,~0.046]$, $[-0.055,~0.060]$,
$[-0.073,~0.110]$ and the average values of absolute deviations
are $0.014$, $0.025$, $0.042$ for $m_{\tilde{\nu}} = 100$, $250$,
$500$ GeV, respectively. The deviations of $K$-factor from the
central value at the LHC are in the ranges of $[-0.057,~0.036]$,
$[-0.044,~0.027]$, $[-0.057,~0.025]$, and the average values of
absolute deviations are $0.018$, $0.016$, $0.019$ for
$m_{\tilde{\nu}} = 100$, $250$, $500$ GeV respectively. The
relative errors of $K$-factor due to the PDF(defined as
$\delta\equiv \frac{K_{max}-K_{min}}{K_{central}}$) for
$m_{\tilde{\nu}}=100$ GeV, $250$ GeV, $500$ GeV, are
$6.0\%(5.8\%)$, $7.8\%(5.0\%)$ and $14.2\%(5.9\%)$ at the
Tevatron(LHC), separately.

\begin{table}
\begin{center}
\begin{tabular}{l|ll|ll|ll} \hline
CTEQ6 & $m_{\tilde{\nu}}=100$ GeV & &  $m_{\tilde{\nu}}=250$ GeV & &  $m_{\tilde{\nu}}=500$ GeV \\
     & $K_{Tevatron}$ & $K_{LHC}$  & $K_{Tevatron}$ &
     $K_{LHC}$  & $K_{Tevatron}$ & $K_{LHC}$ \\ \hline
\hline
 {\bf 6.1M} & 1.643 & 1.614&1.471 &1.418 &1.290 &1.379 \\
 {\bf 201}  & 1.610 & 1.576&1.458 &1.379 &1.285 &1.335\\
 {\bf 202}  & 1.672 & 1.631&1.492 &1.432 &1.293 &1.386\\
 {\bf 203}  & 1.643 & 1.582&1.505 &1.397 &1.339 &1.356 \\
 {\bf 204}  & 1.638 & 1.629&1.444 &1.419 &1.239 &1.366\\
 {\bf 205}  & 1.632 & 1.607&1.505 &1.400 &1.373 &1.349 \\
 {\bf 206}  & 1.648 & 1.601&1.448 &1.411 &1.217 &1.374\\
 {\bf 207}  & 1.590 & 1.591&1.416 &1.376 &1.233 &1.320\\
 {\bf 208}  & 1.689 & 1.616&1.531 &1.432 &1.345 &1.403\\
 {\bf 209}  & 1.625 & 1.565&1.454 &1.381 &1.268 &1.348 \\
 {\bf 210}  & 1.657 & 1.651&1.499 &1.433 &1.313 &1.375 \\
 {\bf 211}  & 1.640 & 1.620&1.479 &1.411 &1.306 &1.358 \\
 {\bf 212}  & 1.644 & 1.591&1.473 &1.403 &1.275 &1.359 \\
 {\bf 213}  & 1.645 & 1.608&1.480 &1.410 &1.304 &1.366 \\
 {\bf 214}  & 1.639 & 1.601&1.473 &1.403 &1.275 &1.355 \\
 {\bf 215}  & 1.638 & 1.596&1.468 &1.396 &1.227 &1.353 \\
 {\bf 216}  & 1.634 & 1.602&1.479 &1.407 &1.351 &1.367 \\
 {\bf 217}  & 1.642 & 1.604&1.473 &1.414 &1.357 &1.377 \\
 {\bf 218}  & 1.629 & 1.596&1.474 &1.391 &1.233 &1.336 \\
 {\bf 219}  & 1.684 & 1.635&1.502 &1.444 &1.320 &1.400 \\
 {\bf 220}  & 1.601 &1.577 &1.451 &1.373 &1.268 &1.324 \\
 {\bf 221}  & 1.634 &1.618 &1.482 &1.416 &1.292 &1.362 \\
 {\bf 222}  & 1.638 &1.601 &1.478 &1.396 &1.308 &1.351 \\
 {\bf 223}  & 1.666 &1.623 &1.495 &1.427 &1.319 &1.384 \\
 {\bf 224}  & 1.655 &1.620 &1.485 &1.420 &1.280 &1.375\\
 {\bf 225}  & 1.668 &1.619 &1.511 &1.425 &1.338 &1.390 \\
 {\bf 226}  & 1.651 &1.620 &1.479 &1.421 &1.285 &1.370 \\
 {\bf 227}  & 1.634 &1.592 &1.512 &1.400 &1.378 &1.366 \\
 {\bf 228}  & 1.645 &1.590 &1.527 &1.393 &1.392 &1.351 \\
 {\bf 229}  & 1.643 &1.618 &1.485 &1.413 &1.324 &1.362 \\
 {\bf 230}  & 1.630 &1.557 &1.496 &1.380 &1.306 &1.350 \\
 {\bf 231}  & 1.646 &1.583 &1.488 &1.397 &1.296 &1.360 \\
 {\bf 232}  & 1.652 &1.620 &1.485 &1.423 &1.298 &1.374 \\
 {\bf 233}  & 1.665 &1.627 &1.492 &1.431 &1.298 &1.380 \\
 {\bf 234}  & 1.666 &1.627 &1.490 &1.430 &1.294 &1.388 \\
 {\bf 235}  & 1.648 &1.595 &1.526 &1.403 &1.396 &1.363 \\
 {\bf 236}  & 1.639 &1.588 &1.529 &1.399 &1.401 &1.354 \\
 {\bf 237}  & 1.637 &1.586 &1.526 &1.398 &1.400 &1.359\\
 {\bf 238}  & 1.647 &1.595 &1.529 &1.399 &1.395 &1.360 \\
 {\bf 239}  & 1.655 &1.606 &1.504 &1.409 &1.339 &1.372 \\
 {\bf 240}  & 1.656 &1.603 &1.513 &1.406 &1.351 &1.366 \\
 \hline
\end{tabular}
\caption{Full set of $K$-factor predictions for the CTEQ family of
PDFs for $m_{\tilde{\nu}}=100$, $250$, $500$ GeV at the Tevatron
and the LHC.}
\end{center}
\end{table}

\par
Considering soft-gluon resummation effects to all the orders in
$\alpha_s$ of leading logarithm, we present the distributions of
the differential cross sections($d\sigma^{QCD}/dq_T$ and
$d\sigma^{resum}/dp_T$) for the processes $p\bar{p}/pp \to
e^+\mu^-+X$ versus the transverse momentum $q_T$ with
$m_{\tilde{\nu}}=250$ GeV and $500$ GeV in Figs.5(a) and 5(b),
where $q_T$ is defined as $q_T^2=(\vec{p}_{eT}+\vec{p}_{\mu
T})^2$. Figure 5(a) is for the process $p\bar p \to e\mu + X$ at
the Tevatron and Figure 5(b) is for the process $pp \to e\mu + X$
at the LHC.
\begin{figure}[htp]
\includegraphics[scale=0.7, bb=0 0 280 240]{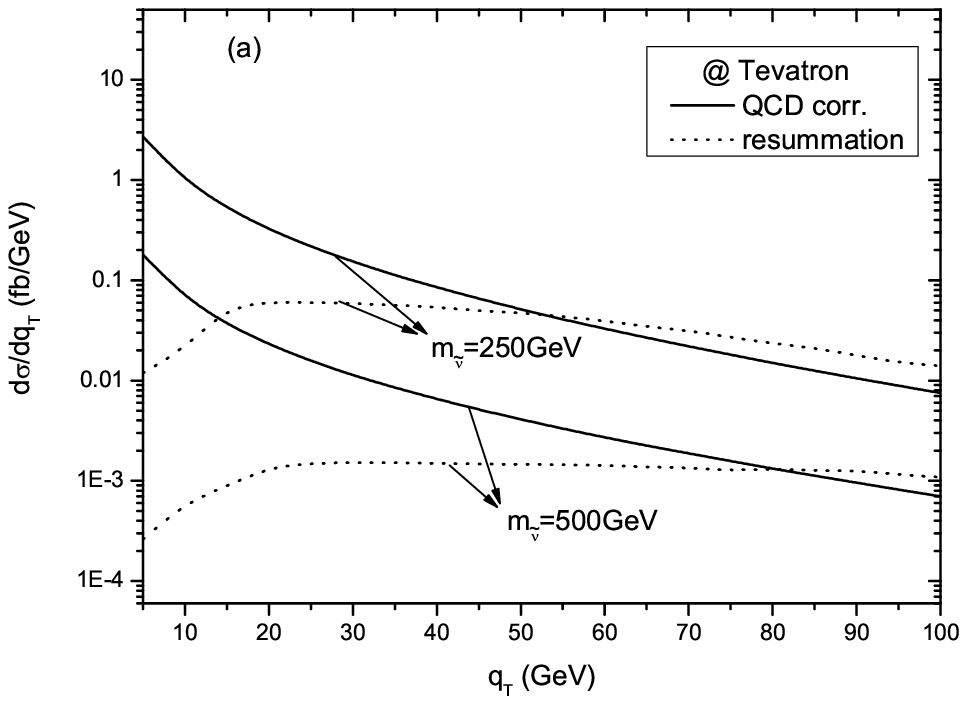}
\includegraphics[scale=0.7, bb=0 0 280 240]{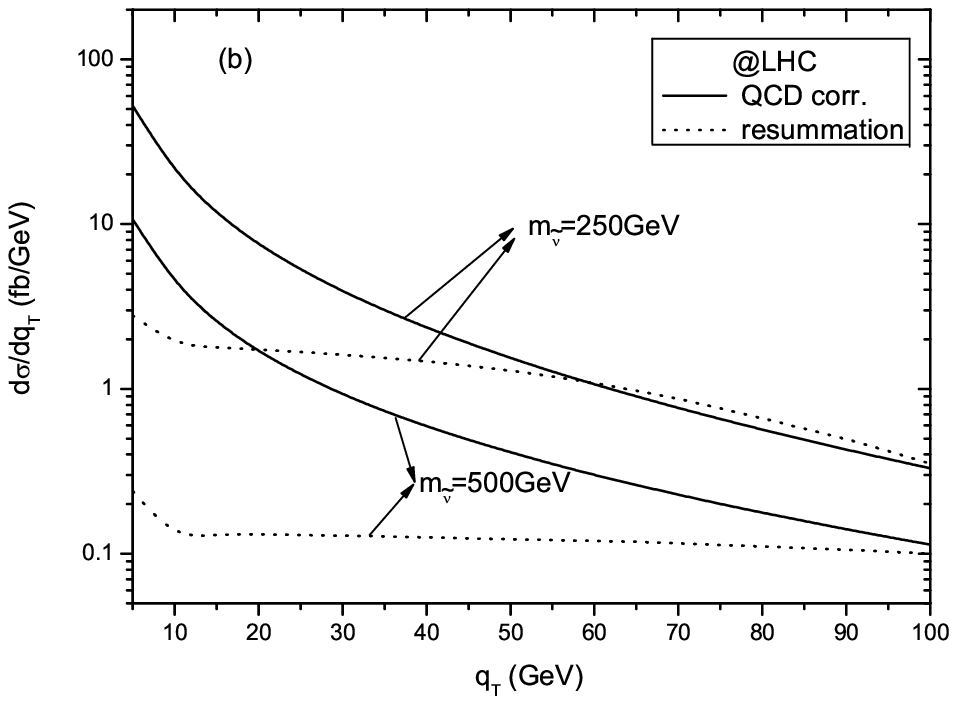}
\caption{Distributions of the transverse momentum of final
$e\mu$-pair $q_T$, which is defined as
$q_T^2=(\vec{p}_{eT}+\vec{p}_{\mu T})^2$. (a) for the Tevatron,
and (b) for the LHC. }
\end{figure}

\par
In summary, our numerical results demonstrate that the QCD
corrections to single sneutrino production cannot directly be
applied to the study of high-mass RPV $e\mu$ pair production. The
$K$-factors of the processes $p\bar p/pp \to e\mu + X$ vary in the
ranges of $[1.182,~1.643]$ and $[1.335,1.614]$ at the Tevatron and
the LHC separately, and the relative errors of $K$-factor are
found to be less than $3\%$($3.1\%$) due to $\mu_f$, and
$14.2\%$($5.9\%$) due to PDF at the Tevatron(LHC) respectively in
our investigating parameter space. We also present the
distributions of the transverse momentum of final $e\mu$-pair by
resummating the logarithmically-enhanced terms for soft gluon as a
reference for future experimental analysis.

\vskip 5mm
\par
\noindent{\large\bf Acknowledgments:} This work was supported in
part by the National Natural Science Foundation of China, the
Education Ministry of China and a special fund sponsored by
Chinese Academy of Sciences.

\end{document}